\pdfoutput=1
\documentclass[
reprint,
bibnotes,
amsmath,amssymb,
]{revtex4-1}

\usepackage{graphicx}% Include figure files
\usepackage{dcolumn}% Align table columns on decimal point
\usepackage{bm}% bold math
\usepackage{setspace}

\usepackage{color}

\begin{document}
\newcommand{\BFA}{BaFe$_{2}$As$_{2}$}
\newcommand{\BKFA}{Ba$_{1-x}$K$_{x}$Fe$_2$As$_2$}
\newcommand{\BFCA}{Ba(Fe$_{1-x}$Co$_{x}$)$_2$As$_2$}
\newcommand{\TCS}{ThCr$_{2}$Si$_{2}$}
\newcommand{\LFAOF}{LaFeAs(O$_{1-x}$F$_x$)}
\newcommand{\TT}{$^{\circ}~2\theta$}
\newcommand{\MB}{$^{57}$Fe-Mö\"ossbauer}
%\preprint{APS/123-QED}

\title{Microscopic co-existence of superconductivity and magnetism in Ba$_{1-x}$K$_{x}$Fe$_2$As$_2$}

\author{Erwin Wiesenmayer$^1$}
\author{Hubertus Luetkens$^2$}
%\email{hubertus.luetkens@psi.ch}
\author{Gwendolyne Pascua$^2$}
\author{Rustem Khasanov$^2$}
\author{Alex Amato$^2$}
\author{Heidi Potts$^3$}
\author{Benjamin Banusch$^3$}
\author{Hans-Henning Klauss$^4$}
\author{Dirk Johrendt$^1$}
\email{johrendt@lmu.de}
\email{hubertus.luetkens@psi.ch}
\affiliation{
{$^1$}{Department Chemie, Ludwig-Maximilians-Universit\"{a}t M\"{u}nchen, D-81377 M\"{u}nchen, Germany}\\
{$^2$}{Labor f\"{u}r Myonenspinspektroskopie, Paul Scherrer
Institute, CH-5232 Villigen PSI, Switzerland} \\
{$^3$}{Swiss Nanoscience Institute (SNI), Universit\"{a}t Basel,
CH-4056 Basel, Switzerland}\\
{$^4$}{Institut f\"ur Festk\"orperphysik, TU Dresden, D--01069
Dresden, Germany}
 }

\date{\today}% It is always \today, today,

\begin{abstract}
It is widely believed that, in contrast to its electron doped counterparts, the hole doped compound \BKFA~exhibits a mesoscopic phase separation of magnetism and superconductivity in the underdoped region of the phase diagram. Here, we report a combined high-resolution x-ray powder diffraction and volume sensitive muon spin rotation study of underdoped \BKFA~($0 \leq x \leq 0.25)$ showing that this paradigm is wrong. Instead we find a microscopic coexistence of the two forms of order. A competition of magnetism and superconductivity is evident from a significant reduction of the magnetic moment and a concomitant decrease of the magneto-elastically coupled orthorhombic lattice distortion below the superconducting phase transition.

\end{abstract}


\maketitle

%\tableofcontents

%\section{\label{intro}Introduction}

% Introduction

The interplay of structural, magnetic and superconducting order parameters is one of the most intriguing aspects in iron based superconductors. In the LaFeAsO (1111) and \BFA~(122) families, superconductivity (SC) evolves from non-superconducting parent compounds with tetragonal crystal structures that are subject to tiny orthorhombic lattice distortions below certain temperatures ($T_s$). Static long-range antiferromagnetic (AF) ordering emerges at N\'{e}el temperatures ($T_N$) well below $T_s$ in LaFeAsO,\cite{Cruz-Neutrons} but very close to $T_s$ in \BFA.\cite{BFA} The structural and magnetic transitions of the parent compounds are suppressed and finally eliminated by doping of the FeAs layers by electrons or holes, and superconductivity emerges at certain doping levels.\cite{Zinth-2011} With respect to the origin of unconventional superconductivity, the possible coexistence of magnetic and superconducting phases in the underdoped areas of the phase diagrams is of considerable interest. But the coupling of structural, magnetic and superconducting order parameters relies on microscopic phase coexistence that is often difficult to distinguish from mesoscopic phase separation.
%due to chemical inhomogeneity.
In the 122-family, microscopic co-existence of these orders is generally accepted for the electron-doped compounds \BFCA, while conflicting reports exist for the hole doped compounds \BKFA.

Co-existence of the orthorhombic structure with SC has first been suggested for \BKFA~up to $x \approx$ 0.2 by x-ray powder diffraction,\cite{Rotter-Angewandte} while neutron diffraction experiments additionally showed long-range AF ordering up to $x \approx$ 0.3. \cite{Chen-BKFA} Diffraction methods however only provide the mean structural information on a rather long spatial scale, and cannot supply conclusive information regarding phase separation. \MB~spectroscopy as a local probe indicated microscopic co-existence,\cite{rotter-2009} but other local probes such as $\mu$SR\cite{Aczel-2008,Goko-2009,Keimer-2009} and NMR\cite{Julien-2009} showed phase separation with non-magnetic superconducting volume fractions between 25 and 40\%. However, most of these experiments were conducted with almost optimally doped \BKFA~single crystals, despite still no growth method yields homogeneous crystals of this compound. Nevertheless, these studies manifested the paradigm that underdoped \BKFA~exhibits mesoscopic phase separation.

In contrast to these scattered results, studies with cobalt-doped \BFCA~yielded convincing evidence for microscopic co-existence.\cite{Pratt-2009} Moreover, competing order parameters became obvious by the concomitant reduction of the orthorhombic lattice distortion $\delta = \frac{a-b}{a+b}$ and magnetic moment $\mu_{\rm{Fe}}$ when crossing the critical temperature.\cite{Nandi-2010} This microscopic co-existence supports $s^\pm$ symmetry of the superconducting order parameter\cite{Fernandes-2010,Fernandes-2010-2} and gives strong evidence for unconventional superconductivity in iron arsenides.

Considering this generally accepted situation for \BFCA, it is particularly important to clarify the intrinsic behavior of \BKFA, also because cobalt-doping causes additional disorder in the (Fe$_{1-x}$Co$_{x}$)-layers, while potassium-doping hardly affects the FeAs-layers. Thus, if both orders co-exist microscopically in \BKFA, we rather observe the behavior of the clean superconducting FeAs-layer. Indeed, a recent neutron diffraction study with polycrystalline material supports early suggestions about microscopic co-existence \cite{Avci-2011}, but gives no conclusive proof regarding microscopic co-existence, because elastic neutron scattering as a bulk probe is principally unable to distinct whether the magnetic volume fraction or the magnetic moment at the iron site decreases.

In this letter, we report a combined high-resolution x-ray diffraction and muon spin rotation ($\mu$SR) study with underdoped \BKFA~($x$=0, 0.19, 0.23, 0.25). We unambiguously show the homogeneous co-existence of the superconducting and antiferromagnetic phase and the competition of the respective order parameters.

% Synthesis

\begin{figure}[h]
\includegraphics[width=1.0\linewidth] {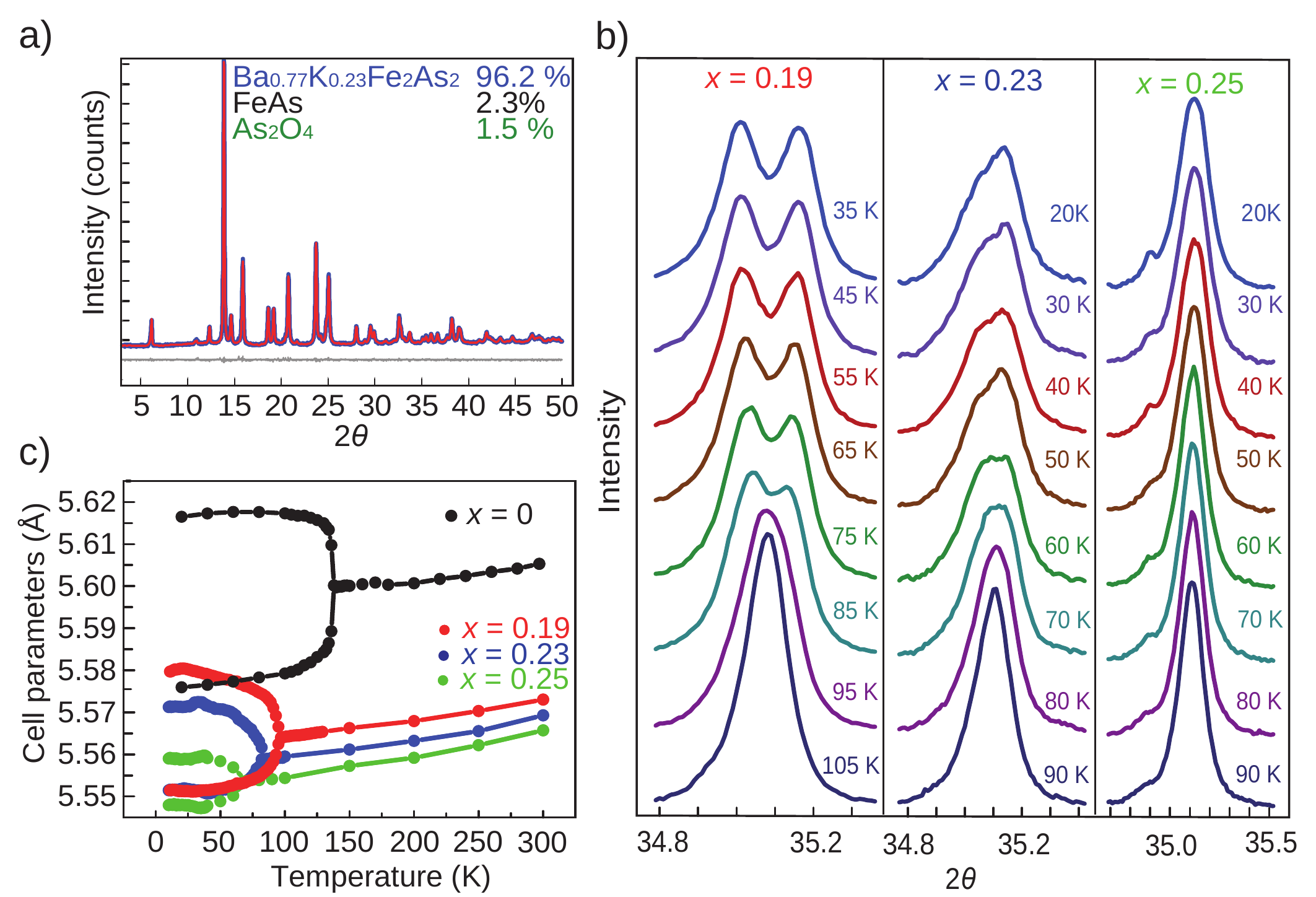}
\caption{\label{fig:XRD} a) X-ray powder diffraction pattern
(blue) and Rietveld fit (red) of \BKFA~($x=0.23$). b) Temperature
dependence of the (112) reflections of for $x=$~0.19, 0.23, and
0.25. c) a- and b-axis cell parameters as a function of
temperature.}
\end{figure}

Polycrystalline samples of \BKFA~were synthesized by heating stoichiometric mixtures of the elements (purities $> 99.9 \%$) in alumina crucibles sealed in silica tubes under purified argon. In order to avoid potassium evaporation, alumina inlays were used. The mixtures were heated to 873~K  (50~K/h) and kept at this temperature for 15~h. The products were homogenized in the crucible, enclosed in a silica tube and annealed at 923~K for 15~h. Finally the powders were cold pressed into pellets, sintered for 20~h at 1023~K and cooled to room temperature by switching off the furnaces. Lattice parameters were obtained by temperature-dependent x-ray powder diffraction (Co,Cu,Mo-$K_{\alpha1}$-radiation) and Rietveld refinements using the TOPAS package \cite{TOPAS}. Fig.~\ref{fig:XRD}~a) shows a typical XRD pattern. Only traces of impurity phases were detected. Ba:K ratios was examined by refinement of the site occupancy parameters and cross checked by ICP-AAS chemical analysis. $\mu$SR measurements have been performed using the GPS and Dolly spectrometers located at the $\pi$M3 and $\pi$E1 beamlines of the Swiss Muon Source at the Paul Scherrer Institut, Switzerland. The data have been analyzed using the MUSRFIT package \cite{Suter-2011}.

X-ray powder patterns of the samples revealed the known structural phase transitions from tetragonal to orthorhombic symmetry. In agreement with our earlier studies \cite{Rotter-Angewandte}, also Ref.\cite{Avci-2011} showed that the orthorhombic distortion depends on the potassium concentration and is finally absent if $x \geq$ 0.3. Fig.\ref{fig:XRD}~b) shows the temperature dependency of the (112) reflections. While the clear splitting, or at least broadening of the peak is visible at $x$ = 0.19 and 0.23, it is apparently absent at $x$ = 0.25. However, a closer inspection reveals the onset of peak broadening below $\approx$70~K also in this case. From this we obtained the tetragonal to orthorhombic transition temperatures $T_{s}=140$~K, 98~K, 84~K and 70~K for $x= 0$, 0.19, 0.23, and 0.25 respectively. The lattice parameters obtained from Rietveld-refinements are shown in Fig.\ref{fig:XRD}~c). It is obvious that potassium doping of \BFA~reduces the transition temperature $T_s$ and also the extent of the lattice parameter splitting, which is still visible at $x$ = 0.25 where $T_c$ is already 32.6 K.

% CO-EXISTENCE
%%%%%%%%%%%%%%

The main goal of this study is to clarify how magnetism and superconductivity coexist in the underdoped region of the \BKFA~phase diagram. For this reason x-ray, AC-susceptibility, and $\mu$SR measurements have been performed on the very same samples to investigate their superconducting and magnetic properties, respectively.

% AC-Susceptibilty

\begin{figure}[h]
\includegraphics[width=1.0\linewidth] {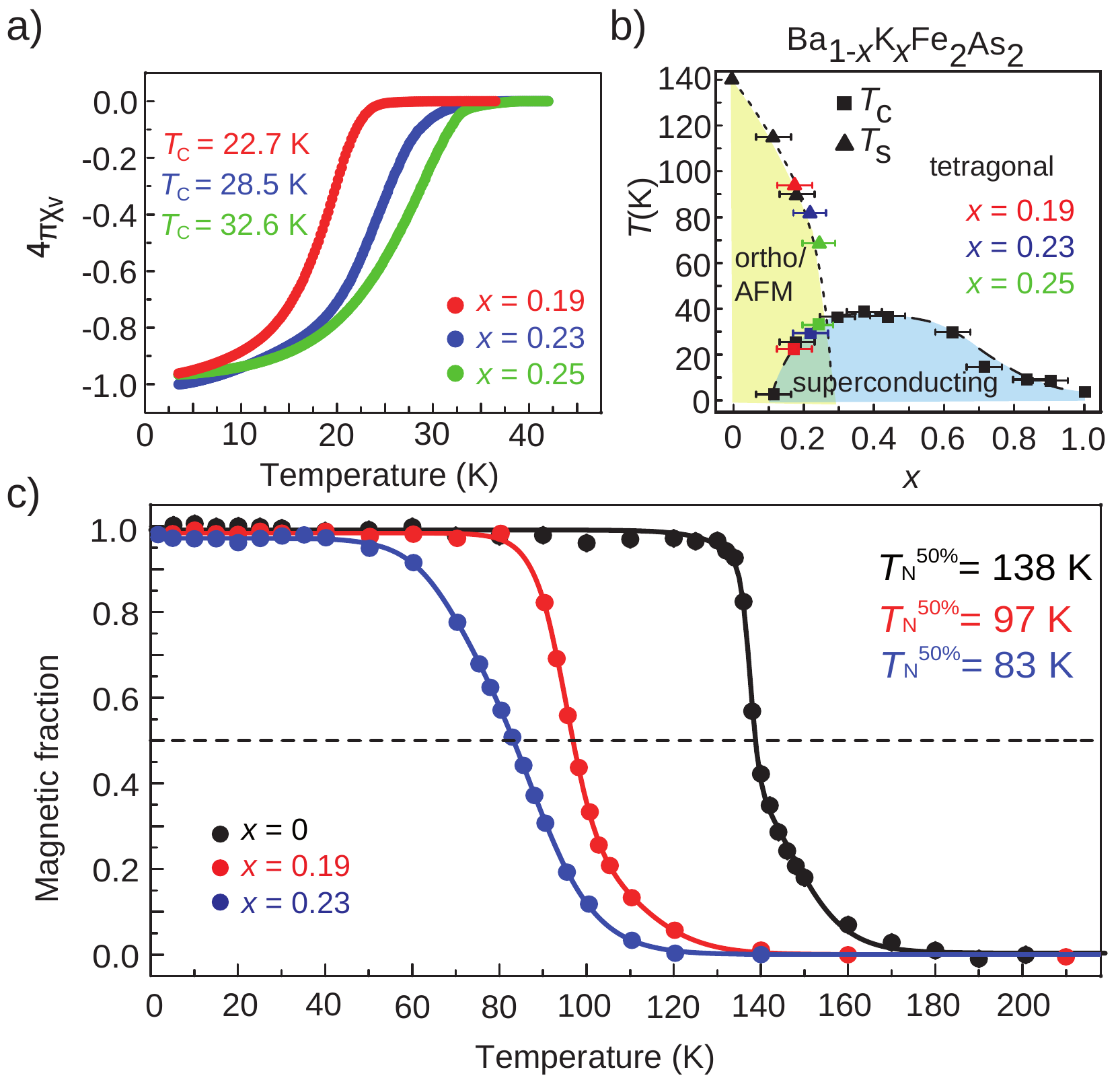}
\caption{\label{fig:Transitions} a) Magnetic susceptibility measurement of \BKFA~with $x$~= 0.19, 0.23, and 0.25 showing $\approx$100\% diamagnetic shielding. b) Structural, magnetic and superconducting phase diagram of \BKFA. c) Magnetic volume fraction as a function of temperature obtained from wTF-$\mu$SR measurements.}
\end{figure}

The AC-susceptibility of finely ground powder samples were measured between 3.4~K and 45~ K at 8~Oe and 1.333~kHz. Diamagnetic signals were detected below $T_c = 22.7$~K ($x$= 0.19),  28.5~K ($x$~= 0.23) and 32.6~K ($x$~= 0.25) as shown in Fig.~\ref{fig:Transitions}~a). The superconducting volume fractions of all samples are close to 100\% and prove bulk superconductivity.

Muon spin rotation measurements in a weak transverse field (wTF-$\mu$SR) provide an easy means to measure the magnetic volume fraction.
%In the paramagnetic state of the specimen, the muon spins precess about the weak externally applied magnetic field $B_\mathrm{ext}$ with a Larmor frequency $\omega = \gamma_\mu B_\mathrm{ext}$ with $\gamma_\mu$ being the gyromagnetic ratio of the muon. The amplitude of that precession signal is a measure for the paramagnetic volume fraction. In the magnetic state, the internal magnetic fields overcome by far the externally applied field so that only muons exhibiting a non-magnetic environment contribute to the precession at $\omega = \gamma_\mu B_\mathrm{ext}$.
In Fig.~\ref{fig:Transitions}~c) the magnetic volume fractions obtained by such measurements in $B_\mathrm{ext}= 50$~Oe are shown for various \BKFA~samples ($x=0.0$, 0.19 and 0.23) as a function of temperature. For all samples a transition to a magnetic state is observed. From this the magnetic transition temperature where 50\%  of the volume is magnetic has been determined to $T_\mathrm{N} = 138$~K, 97 K and 83 K for the three samples respectively. The magnetic volume fraction reaches 100\% for all three samples and, most remarkable, does not change below the superconducting $T_c$. Therefore, these results, together with the 100\% superconducting shielding signal observed in the AC-susceptibility measurements, prove the microscopic coexistence of magnetism and superconductivity in the orthorhombic phase of our samples. The structural, magnetic and superconducting transition temperatures are compiled in the phase diagram depicted in Fig.~\ref{fig:Transitions}~b).

%Low temperature structure measurement

\begin{figure}[h]
\includegraphics[width=0.7\linewidth] {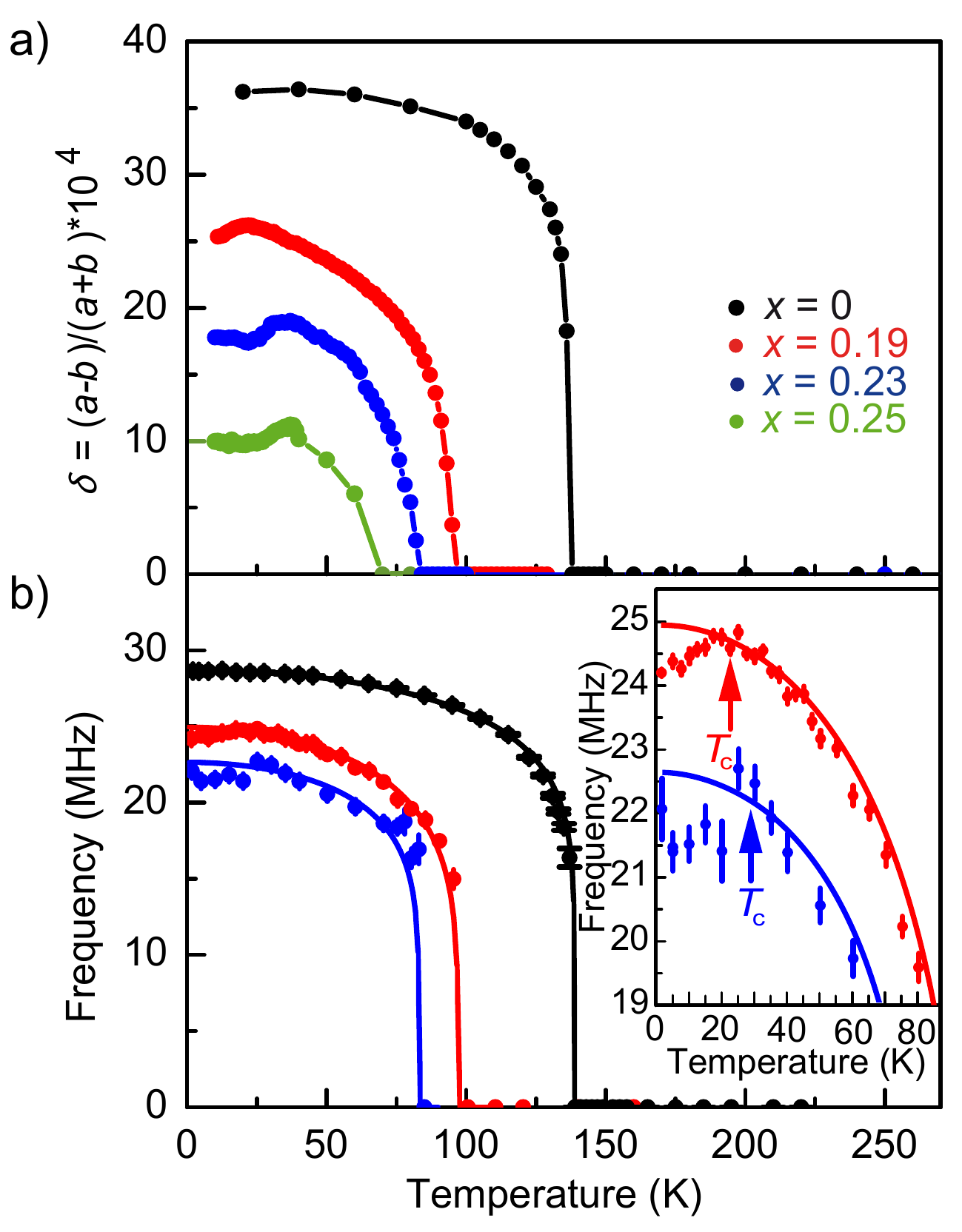}
\caption{\label{fig:OrderParameter} Orthorhombicity parameter $\delta$ and ZF-$\mu$SR frequency (magnetic order parameter) of \BKFA~as a function of temperature.}
\end{figure}

The orthorhombic distortion in terms of the structural order parameter $\delta$ is shown in Fig.~\ref{fig:OrderParameter}~a). In the $x$ = 0.19 sample, $\delta$ achieves a clear maximum $\delta_{\rm{max}} \approx 27\times10^{-4}$ at the superconducting transition temperature close to 23~K and then decreases to tower temperatures. Higher potassium concentrations further decrease $T_s$ to 84~K while $\delta_{\rm{max}} \approx 20\times10^{-4}$ again coincides with $T_c$ at 28.5~K ($x$ = 0.23). This trend continues to $x$ = 0.25 with $T_s \approx 70~K$, $T_c$ = 32.6 K and $\delta_{\rm{max}} \approx 13\times10^{-4}$. This behavior is similar to \BFCA,\cite{Nandi-2010} however, we do not observe the further linear decrease of $\delta$ at lower temperatures back to a quasi-tetragonal structure, but rather saturation of $\delta$. Also in contrast to the Co-doped material, we find that the effect becomes smaller with increasing potassium concentrations $x$. The reason for that is not yet clear. We suggest that the stronger effect in the case of Co-doping may be connected with the fact, that magnetic ordering is weakened not only by the electron-doping, but additionally by the disorder that is introduced by the cobalt-atoms at the iron sites. Thus the competition of superconductivity and antiferromagnetism for the same electrons may affect the (Fe$_{1-x}$Co$_{x}$)As layers more efficiently than the clean FeAs-layers in the K-doped material.

\begin{figure}[h]
\includegraphics[width=0.70\linewidth] {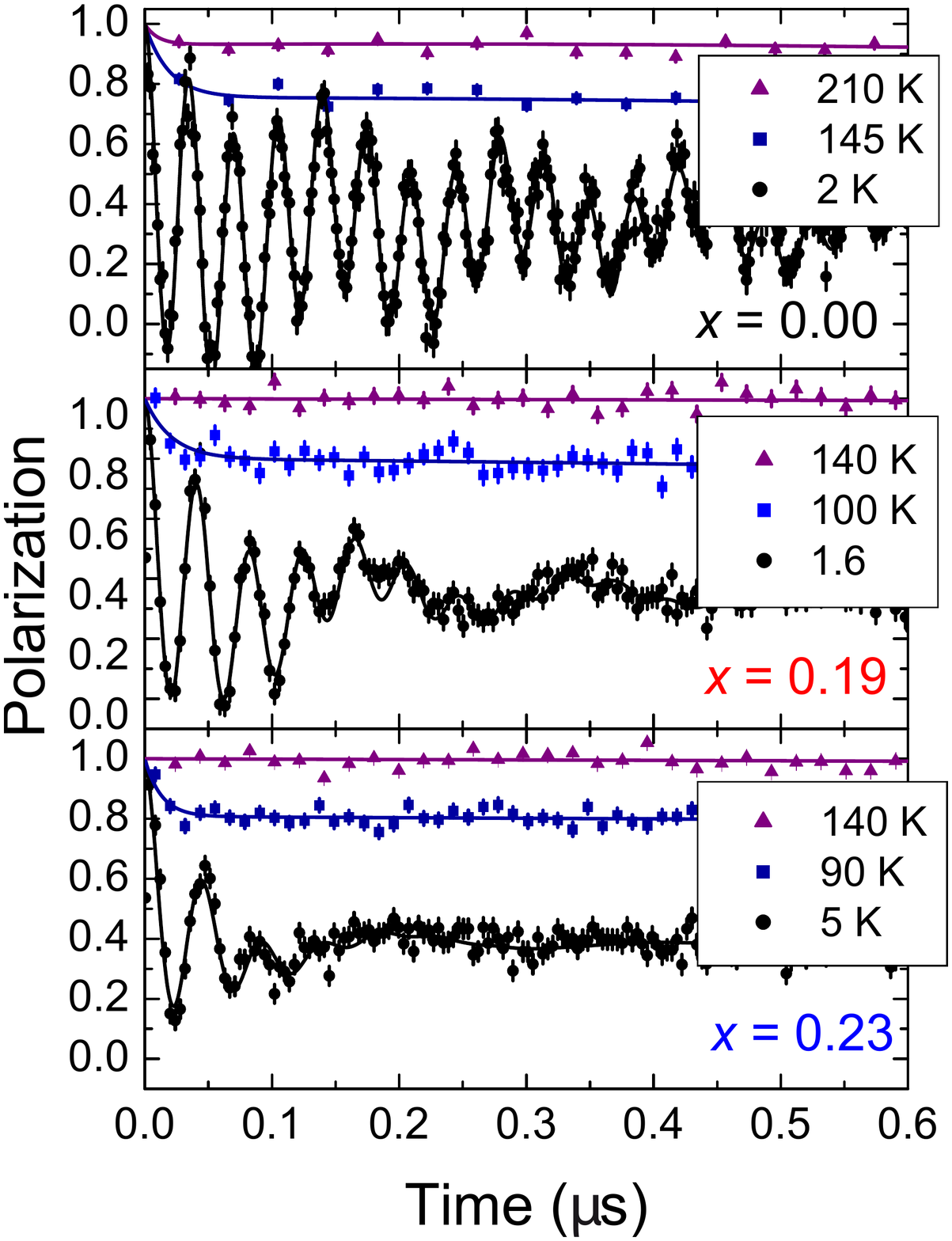}
\caption{\label{fig:musr} Zero field $\mu$SR spectra for
characteristic temperatures (above, at, and below the transition)
for \BKFA~powder samples with $x$~= 0.0, 0.19, and 0.23.}
\end{figure}

To elucidate further the magnetic properties of \BKFA ~ zero field (ZF) $\mu$SR measurements have been performed. The ZF-$\mu$SR spectra shown in Fig.~\ref{fig:musr} exhibit well defined muon spin precessions at below $T_\mathrm{N}$ even for the $x=0.23$ sample indicating a long-range ordered magnetic phase.

As already observed in other Fe-based superconductors \cite{Klauss-2008,Jesche-2008,Maeter-2009,Goko-2009} the ZF spectra are composed of two distinct precession frequencies which has been interpreted as two magnetically inequivalent muon stopping sites in the structure. The data can be well fitted with a damped cosine functional form indicating a commensurate magnetic structure.\cite{Yaouanc-2011} Please note, that the in under-doped samples of the related \BFCA~ family only a strongly over-damped oscillation can be observed \cite{Bernhard-2009,Marsik-2010}. This indicates that the doping in the Ba layer causes considerably less disorder into the magnetic system. Another difference is that in the Co-doped systems $\mu$SR spectra consistent with incommensurate order have been found. ZF-$\mu$SR allows to precisely determine the temperature dependence of the magnetic order parameter (Fe moment) which is proportional to the measured $\mu$SR frequency. The observed $\mu$SR frequency is shown in Fig.~\ref{fig:OrderParameter}~b) together with the orthorhombicity parameter $\delta = \frac{a-b}{a+b}$ deduced from the XRD measurements.

The magnetic order parameter shown in Fig.~\ref{fig:OrderParameter}~b) decreases alike the orthorhombicity as a function of potassium doping. Most remarkable however is the decrease of the magnetic order parameter (Fe moment) below the superconducting $T_c$ clearly visible in the inset of Fig.~\ref{fig:OrderParameter}~b). Here we would like to mention that $\mu$SR as a local probe is able to measure the magnetic volume fraction (as shown above) and the size of the ordered moment (via the ZF-$\mu$SR frequency) separately unlike it is done in scattering experiments where the product of both quantities is measured. Taking all data together it is obvious that all investigated samples remain 100\% magnetic, but that the ordered Fe magnetic moment as well as the orthorhombicity decrease below the superconducting $T_c$. In other words, superconductivity and magnetism coexist on a microscopic scale, but compete for the same electrons in the underdoped region of the \BKFA~phase diagram.

% Conclusion

In summary, our results prove the paradigm of phase-separation in \BKFA~wrong. Instead we find compelling evidence of microscopic co-existence of superconductivity with magnetic ordering from combined x-ray and $\mu$SR data. The competition for the same electrons reduces the magnetic moment below $T_c$, while the magnetic fraction remains 100~\% according to volume-sensitive $\mu$SR measurements. The response of the structural and magnetic order parameters at $T_c$ is weaker than in Co-doped \BFCA. Since K-doping introduces no disorder in the superconducting FeAs-layer, we suggest that we rather observe its intrinsic behavior.

\begin{acknowledgments}

Thanks to Marianne Rotter for the low temperature X-ray diffraction measurements and to Marcus Tegel for support with Rietveld refinements. This work was financially supported by the German Research Foundation (DFG) within the priority program SPP1458, project No. JO257/6-1. Part of this work has been performed at the Swiss Muon Source at the Paul Scherrer Institut, Switzerland.

\end{acknowledgments}

\bibliographystyle{apsrev}

%\bibliography{BKFA_ortho}

\end{document}